\documentclass[]{spie}  

\usepackage{amsmath,amsfonts,amssymb}
\usepackage{graphicx}
\usepackage[colorlinks=true, allcolors=blue]{hyperref}
\usepackage{amssymb}
\usepackage{adjustbox}
\usepackage{graphicx}
\usepackage{siunitx}
\usepackage{mathtools}
\usepackage{threeparttable}
\usepackage{float}
\usepackage{booktabs}
\usepackage{tabularx}
\usepackage{makecell}
\newcolumntype{C}[1]{>{\centering\arraybackslash}p{#1}}
\newcolumntype{L}[1]{>{\raggedright\arraybackslash}p{#1}}
\usepackage{multicol}
\usepackage[T1]{fontenc}
\usepackage{csquotes}
\usepackage{url}
\usepackage{subcaption}

\title{Universal Lesion Detection in CT Scans using Neural Network Ensembles}

\author[]{Tarun Mattikalli}
\author[]{Tejas Sudharshan Mathai}
\author[]{Ronald M. Summers}
\affil[]{Imaging Biomarkers and Computer-Aided Diagnosis Laboratory, Radiology and Imaging Sciences, Clinical Center, National Institutes of Health, Bethesda MD, USA}

\authorinfo{Further author information: (Send correspondence to T.S.M.)\\T.S.M.: E-mail: tejas.mathai@nih.gov}

\begin{document} 
\maketitle

\begin{abstract}
In clinical practice, radiologists are reliant on the lesion size when distinguishing metastatic from non-metastatic lesions. A prerequisite for lesion sizing is their detection, as it promotes the downstream assessment of tumor spread. However, lesions vary in their size and appearance in CT scans, and radiologists often miss small lesions during a busy clinical day. To overcome these challenges, we propose the use of state-of-the-art detection neural networks to flag suspicious lesions present in the NIH DeepLesion dataset for sizing. Additionally, we incorporate a bounding box fusion technique to minimize false positives (FP) and improve detection accuracy. Finally, to resemble clinical usage, we constructed an ensemble of the best detection models to localize lesions for sizing with a precision of 65.17\% and sensitivity of 91.67\% at 4 FP per image. Our results improve upon or maintain the performance of current state-of-the-art methods for lesion detection in challenging CT scans.
\end{abstract}




\section{Purpose}
\label{purpose}

The purpose of this work is to design an ensemble-based universal lesion detector composed of state-of-the-art detection networks to localize lesions in CT scans present in the NIH DeepLesion dataset.

\section{Introduction}
\label{intro}  

Localization of lesions allows radiologists to identify those that are enlarged, malignant, and at risk for tumor spread. Currently, Computed Tomography (CT) is the preferred imaging modality \cite{Schwartz2016} for lesion assessment and diagnosis, but other modalities provide complementary information, e.g. Positron Emission Tomography (PET). Moreover, lesions in CT have asymmetrical shapes and diverse appearances. Yet, radiologists rely heavily on the lesion size measurement \cite{Schwartz2016}, which plays a major role in the course of therapy provided to the patients. Lesions in CT scans are sized according to the RECIST guidelines \cite{Schwartz2016}, which constitutes two orthogonal lines representing the long and short axis of the suspected lesion. This guideline can vary among institutions due to the type of CT scanner used (conventional vs spiral), exam protocols, clinical practice, etc. Lesions of all sizes can be suspicious for metastasis, but there are often lesions that shrink and reappear at later patient visits, and radiologists often miss such smaller regressed lesions. Thus, there is a need for automated universal lesion detection in CT scans for cancer burden assessment. 

Several state-of-the-art (SOTA) approaches \cite{Yan2019_mulan,Yan2018_3dce,Tang2019_uldor,Zlocha2019,Xie2021_recistnet,Cai2021_lesionharvester,Yan2020_multData} for universal lesion detection have been proposed. However, most approach the lesion detection problem with a heavy redesign of the underlying neural network \cite{Yan2019_mulan,Yan2018_3dce,Xie2021_recistnet,Tang2019_uldor}, computationally expensive hyperparameter optimization \cite{Zlocha2019} (e.g. refining the anchor box size and aspect ratio), and dataset supplementation \cite{Yan2020_multData,Cai2021_lesionharvester}. In this pilot work, we detect lesions in CT scans comprising the NIH DeepLesion dataset \cite{Yan2018_deeplesion} with SOTA detection networks \cite{Lin2017_retinanet,Kong2019_foveabox,Zhang2021_vfnet}, and improve upon their performance with a simple bounding box fusion technique \cite{Solovyev2021} that significantly reduces the false positive rate. Furthermore, we propose an ensemble of the best detection models, show its clinical applicability for lesion detection, and detail results that improve upon the performance of previously published lesion detection methods.

\begin{figure}[h]

\centering
\begin{subfigure}[b]{0.19\columnwidth}
\vspace*{\fill}
  \centering
  \includegraphics[width=\columnwidth,height=2.55cm]{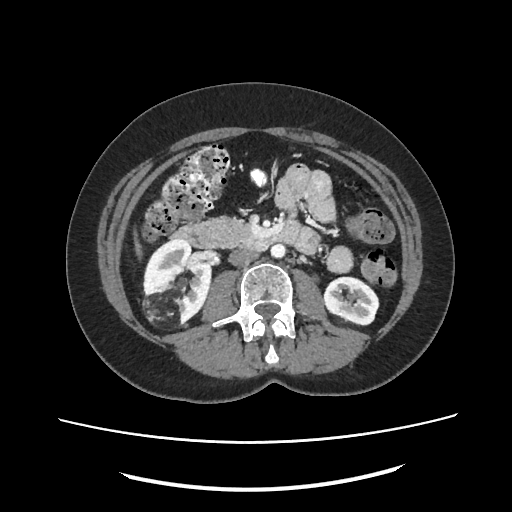}
  \includegraphics[width=\columnwidth,height=2.55cm]{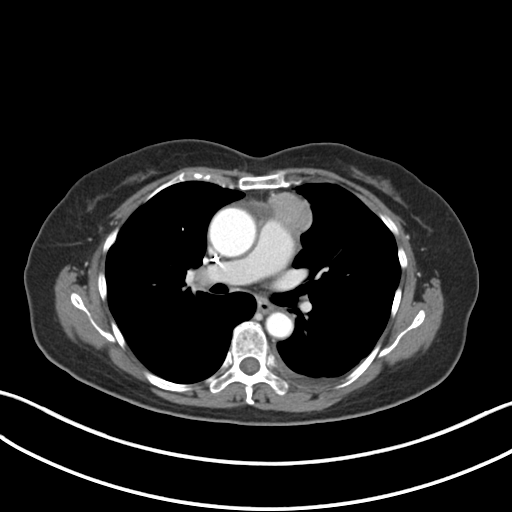}
  \includegraphics[width=\columnwidth,height=2.55cm]{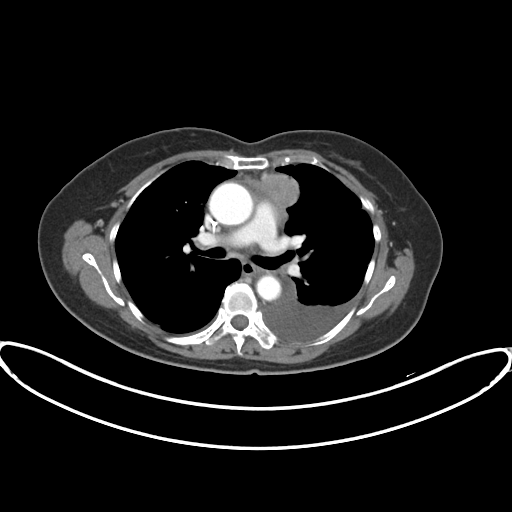}

  \centerline{(a) Original}
\end{subfigure} 
\begin{subfigure}[b]{0.19\columnwidth}
\vspace*{\fill}
  \centering
  \includegraphics[width=\columnwidth,height=2.55cm]{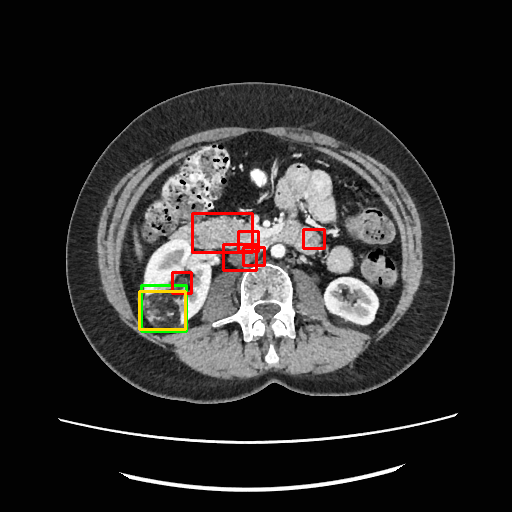}
  \includegraphics[width=\columnwidth,height=2.55cm]{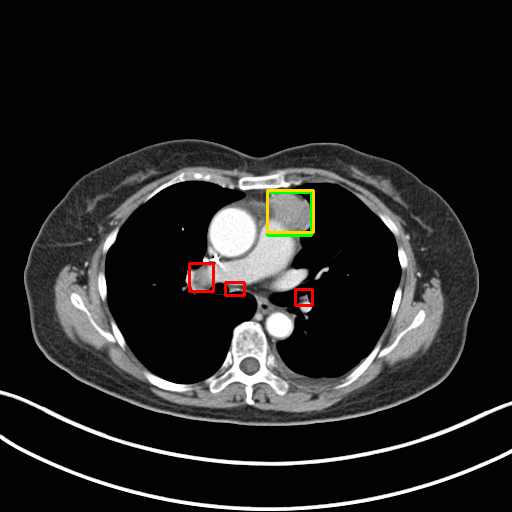}
  \includegraphics[width=\columnwidth,height=2.55cm]{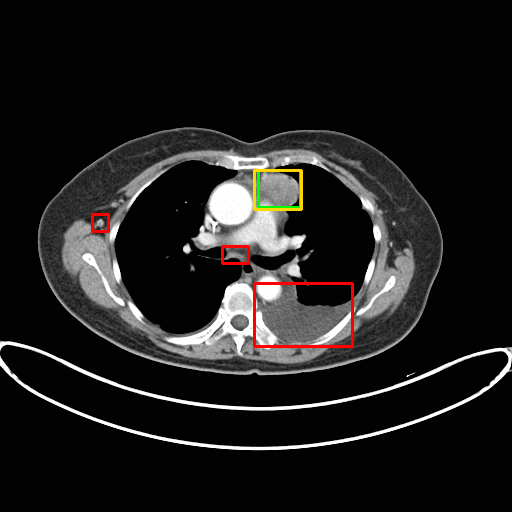}
  \centerline{(b) VFNet}
\end{subfigure} 
\begin{subfigure}[b]{0.19\columnwidth}
\vspace*{\fill}
  \centering
  \includegraphics[width=\columnwidth,height=2.55cm]{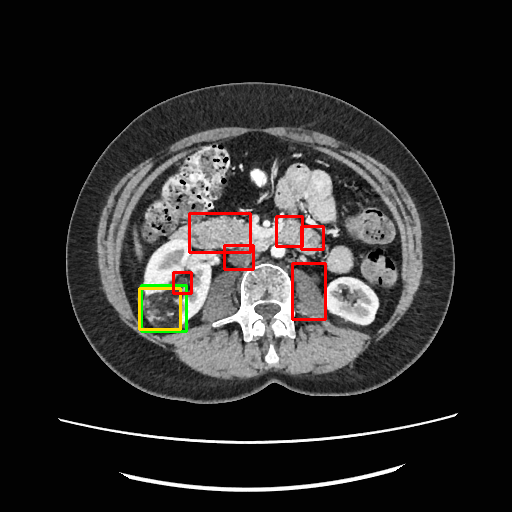}
  \includegraphics[width=\columnwidth,height=2.55cm]{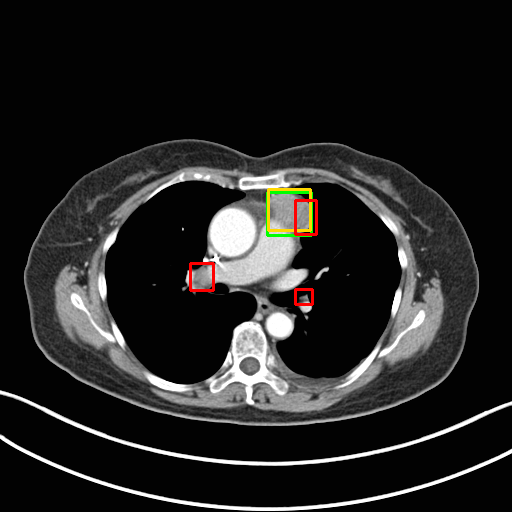}
  \includegraphics[width=\columnwidth,height=2.55cm]{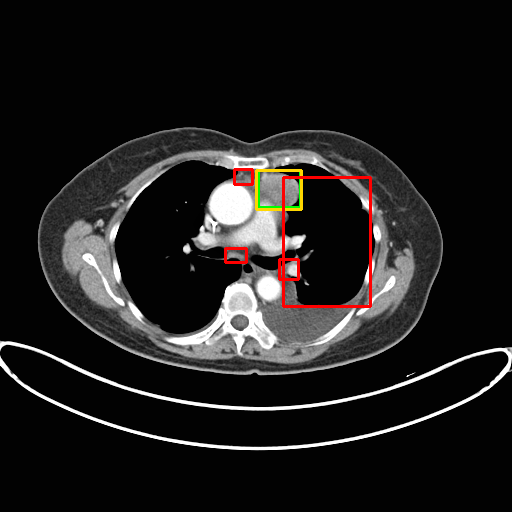}
  \centerline{(c) FoveaBox}
\end{subfigure} 
\begin{subfigure}[b]{0.19\columnwidth}
\vspace*{\fill}
  \centering
  \includegraphics[width=\columnwidth,height=2.55cm]{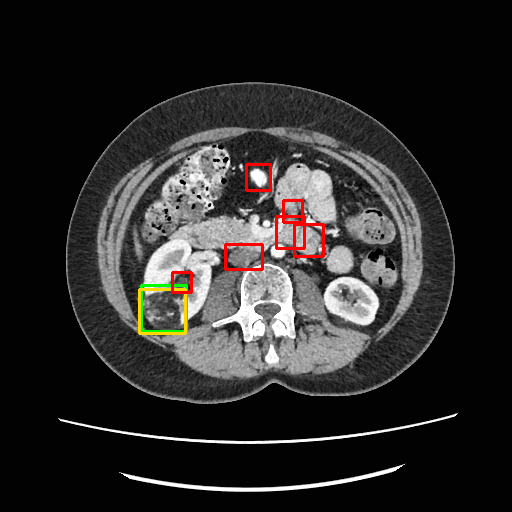}
  \includegraphics[width=\columnwidth,height=2.55cm]{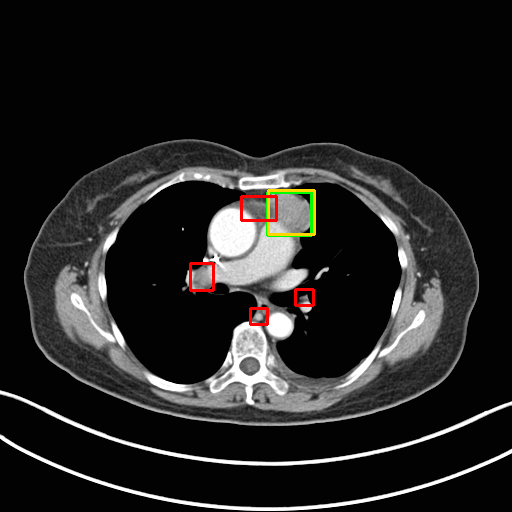}
  \includegraphics[width=\columnwidth,height=2.55cm]{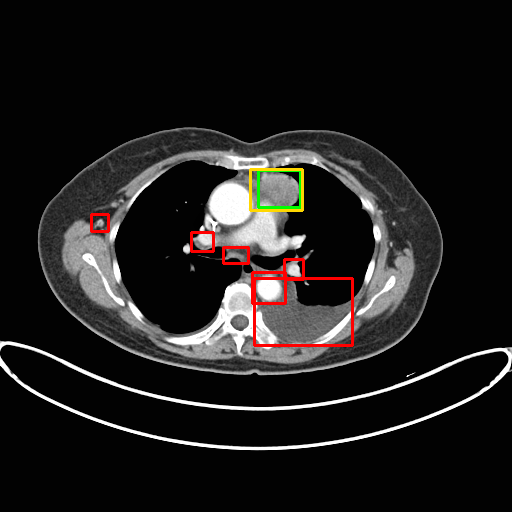}
  \centerline{(d) RetinaNet}
\end{subfigure} 
\begin{subfigure}[b]{0.19\columnwidth}
\vspace*{\fill}
  \centering
  \includegraphics[width=\columnwidth,height=2.55cm]{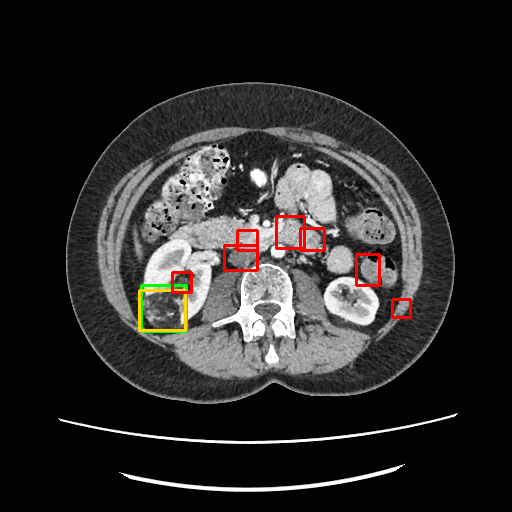}
  \includegraphics[width=\columnwidth,height=2.55cm]{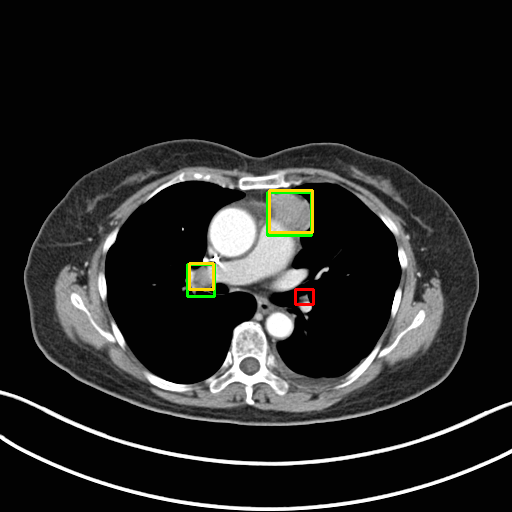}
  \includegraphics[width=\columnwidth,height=2.55cm]{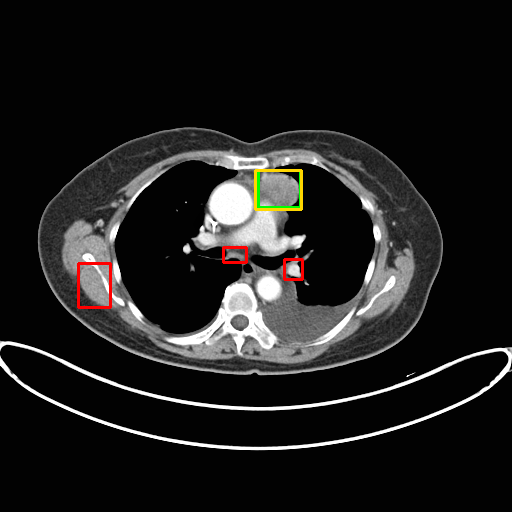}
  \centerline{(e) Ensemble}
\end{subfigure} 
\caption{Column (a) shows example images from the NIH DeepLesion dataset after windowing, but before histogram normalization. (b) - (d) show the lesion detection results of three networks: VFNet, FoveaBox, and RetinaNet. Column (e) shows the detection result of the ensemble after weighted boxes fusion. Green boxes indicate the ground truth, yellow boxes indicate the true positives, and red boxes indicate false positives.}
\label{fig:qual_images}
\end{figure}

\section{Methods}

The goal of this work is to design a method that can universally identify lesions in the NIH DeepLesion dataset, which consists of challenging CT scans. First, we identified the SOTA detection networks that achieve an optimal trade-off between detection sensitivity and localization precision. Next, we employed a bounding box fusion approach to decrease the false positive detection rate, while also promoting generalization. Finally, we ensembled the best models that yielded the optimal performance for the lesion detection task.

\noindent
\textbf{State-of-the-Art Object Detectors.} We quantified the performance of three SOTA one-stage detection networks on the lesion detection task: 1) RetinaNet \cite{Lin2017_retinanet}, 2) FoveaBox \cite{Kong2019_foveabox}, and 3) VFNet \cite{Zhang2021_vfnet}. These detectors directly estimate the class probabilities and bounding box coordinates in a single forward pass. They can further be subdivided into anchor-based (RetinaNet) and anchor-free detectors (FoveaBox and VFNet). RetinaNet \cite{Lin2017_retinanet} overcame the common class imbalance problem plaguing detection tasks by using a focal loss function, along with the Feature Pyramid Network (FPN) to represent objects at different scales. VFNet \cite{Zhang2021_vfnet} combines FCOS (without centerness branch) with an Adaptive Training Sample Selection (ATSS), sets the IoU between the ground truth and the prediction as the classification score, integrates it into a novel IoU-aware Varifocal loss, develops a star-shaped bounding box representation, and refines the box predictions. FoveaBox \cite{Kong2019_foveabox} consists of a backbone to compute features from the input and a fovea head network that estimates the object occurrence possibility through per pixel classification on the backbone's output, and predicts the box at each position in the image that may be potentially covered by an object. We chose the anchor-free detectors in our experiment as they achieve superior results over anchor-based and even two-stage detectors (e.g. Faster RCNN).

\noindent
\textbf{Weighted Boxes Fusion.} Typically, detectors generate multiple bounding box predictions (with confidence scores) of a lesion's location. Predictions become clustered together when an ensemble is created from multiple selected epochs of a single detector or over different object detection networks. Weighted boxes fusion \cite{Solovyev2021} was used to combine the clustered predictions and yield precise detections.

\noindent
\textbf{Ensemble of Best Detection Models.} We hypothesized that a combination of one-stage detectors would promote generalization and boost the detection accuracy in comparison to an individual model. To this end, we ensembled the models (VFNet, RetinaNet, and FoveaBox) to improve the detection capability.

\section{Experiments and Results}

\subsection{Dataset Description}

The NIH DeepLesion dataset \cite{Yan2018_deeplesion} contains 32,120 axial CT scans extracted from 10,594 CT scans, which were obtained after scanning 4,427 unique patients. There are 1-3 lesions per slice, and for context, 30mm of slices above and below the keyslice are provided. RECIST measurements are provided for each lesion along with the 2D bounding box coordinates. For training, the pixel intensities in the original Hounsfield units range were clipped to the corresponding DICOM window provided for each slice (e.g. [-1500, 500]). Then, the images were normalized to the [0,255] range and finally histogram equalized.

\subsection{Implementation}

We use the official splits of the DeepLesion dataset (70\% training, 15\% validation, and 15\% testing) to compare our results directly against prior work \cite{Tang2019_uldor,Yan2018_3dce,Zlocha2019,Xie2021_recistnet}. As radiologists scroll through 1-3 slices of a scan when sizing the lesion extent, we generated 3-slice CT images with the center slice containing the annotated lesion, and used the 3-slice input for training the models using the mmDetection framework \cite{Chen2019_mmdet}. A ResNet50 backbone was used for the RetinaNet, FoveaBox, and VFNet models, and standard data augmentation was performed: random flips, crops, shifts and rotations in the range of [0, 32] pixels and [0, 10] degrees respectively. Batch size and learning rates were set according to the original training schema for each model, and each model was trained for a maximum of 12 epochs. During testing, predictions from the top 5 epochs with lowest validation loss were combined together using weighted box fusion to improve detection performance. In contrast, the models in the final detection ensemble only had 1 epoch with the lowest validation loss. Experiments were run on workstations running Ubuntu 16.04LTS containing Quadro P200X GPUs.

\subsection{Results}

A summary of our results is shown in Table \ref{table_lesion_detection_results}. Amongst the SOTA detection networks, VFNet performed the best with 64\% precision and 82.95\% sensitivity at 4 FP per image. FoveaBox was the next best model with 50.51\% precision and 83.95\% sensitivity followed by RetinaNet. This indicated that RetinaNet had significant issues with detecting LN $\leq$ 10mm \cite{Zlocha2019}. Although the VFNet performance was promising, it was difficult to establish a clear winner amongst the one-stage detectors. As these detectors are computationally less expensive to train, we hypothesized that the combination of one-stage detectors would boost the detection accuracy over VFNet. To this end, we created an ensemble model from VFNet, FoveaBox, and RetinaNet, and obtained a mAP of 65.71\% and a sensitivity of 91.69\% at 4 FP. 

\begin{table}[!h]
\centering\fontsize{9}{12}\selectfont 
\setlength\aboverulesep{0pt}\setlength\belowrulesep{0pt} 
\setlength{\tabcolsep}{7pt} 
\setcellgapes{3pt}\makegapedcells 
\caption{Detection performance of various detectors and our proposed ensemble method. `S" stands for Sensitivity @[0.5, 1, 2, 4, 6, 8, 16] FP. \mbox{--} indicates unavailable metric values.}
\begin{adjustbox}{max width=\textwidth}
\begin{tabular}{@{} c|c|c|c|c|c|c|c|c @{}} 
\toprule
Method                                                          & mAP       & S@0.5     & S@1       & S@2       & S@4       & S@6       & S@8       & S@16 \\
\midrule

3DCE \cite{Yan2018_3dce} (27 slices)                            & --        & 62.48     & 73.37     & 80.70     & 85.65     & --        & 89.09     & 91.06 \\
ULDOR \cite{Tang2019_uldor}                                     & --        & 52.86     & 64.80     & 74.84     & 84.83     & --        & 87.17     & 91.80 \\
Improved RetinaNet \cite{Zlocha2019}                            & --        & 72.15     & 80.07     & 86.40     & 90.77     & --        & 94.09     & 96.32 \\
RECIST-Net \cite{Xie2021_recistnet} (original, 3 slices)        & --        & 74.33     & 81.80     & 87.66     & 90.68     & --        & --        & -- \\

\midrule
RetinaNet \cite{Lin2017_retinanet}                              & 50.01     & 55.79     & 68.76     & 77.57     & 83.86     & 86.15     & 87.51     & 87.51 \\
FoveaBox \cite{Kong2019_foveabox}                               & 50.51     & 55.67     & 67.46     & 77.26     & 83.95     & 86.50     & 87.39     & 87.39 \\
VFNet \cite{Zhang2021_vfnet}                                    & 64.00     & 64.02     & 66.73     & 74.93     & 82.95     & 88.30     & 88.30     & 88.30 \\
Ensemble (VFNet + FoveaBox + RetinaNet)                         & \textbf{65.71}     & 67.78     & 77.77     & 85.75     & \textbf{91.69}     & 92.38     & 92.38     & 92.38 \\
\midrule

Ensemble (SAD $<$  10mm)                                        & 59.77     & 61.71     & 70.83     & 79.68     & 85.05     & 86.35     & 86.92     & 87.01 \\
Ensemble (10mm \mbox{--} 30mm)                                      & 68.24     & 69.68     & 79.12     & 87.14     & 91.99     & 93.32     & 93.89     & 93.89 \\
Ensemble (SAD $\geq$ 30mm)                                      & 60.11    & 66.89     & 80.87     & 88.35     & 91.59     & 92.81     & 93.79     & 93.79 \\
\bottomrule
\end{tabular}
\end{adjustbox}
\label{table_lesion_detection_results}
\end{table}

When compared against 3DCE \cite{Yan2018_3dce}, our recall is 91.69\% vs 85.65\% at 4 FP per image, whereas against ULDOR \cite{Tang2019_uldor}, our sensitivity is 91.69\% vs 84.83\% at 4 FP. Contrasted against the improved RetinaNet \cite{Zlocha2019} and RECIST-Net \cite{Xie2021_recistnet}, our recall was higher by $\sim$1\%. Next, we subdivided our dataset based on the lesion size and evaluated the performance of the ensemble detector. Our ensemble yields a modest performance of 59.77\% mAP and sensitivity of 85.05\% at 4 FP per image when tested on lesions with a SAD of $\leq$10mm. This is comparable to the 88.35\% sensitivity achieved by a highly fine-tuned RetinaNet \cite{Zlocha2019}. On lesions with a SAD between 10mm \mbox{--} 30mm and SAD $\geq$10mm, we attained a mAP/sensitivity of 68.24\%/91.99\% and 60.11\%/91.59\% respectively at 4 FP per image. Our values are consistent with prior work \cite{Zlocha2019,Xie2021_recistnet}, yet we outperform them (sometimes considerably) with a $\geq$1\% increase in sensitivity.

\section{New Work}

We designed a universal lesion detection algorithm to localize lesions that are suspicious for metastasis in the NIH DeepLesion dataset. We first studied the performance of state-of-the-art (SOTA) detection neural networks on the lesion detection task and merged predictions with the weighted boxes fusion technique to reduce the false positive rate. To emulate clinical use, we developed an ensemble of the best performing detection networks to achieve a clinically acceptable detection precision of 65.17$\%$ and sensitivity of 91.69$\%$ at 4 FP per image. Our results outperform previously published lesion detection methods on the challenging NIH DeepLesion CT scans.

\section{Conclusions}

In this work, we quantify the performance of three state-of-the-art one-stage detectors on the challenging task of universal lesion detection. Next, we combined the detectors in an ensemble, and applied Weighted Boxes Fusion to merge the predicted bounding box clusters produced by the ensemble. The ensemble was able to detect lesions better with a precision of 65.71\%  and sensitivity of 91.69\% at 4 FP per image. Our approach outperforms prior work proposed to tackle universal lesion detection on the NIH DeepLesion dataset.

\section{Acknowledgements.} 

This work was supported by the Intramural Research Programs of the NIH Clinical Center and NIH National Library of Medicine. We also thank Jaclyn Burge for the helpful comments and suggestions.

\bibliography{report} 
\bibliographystyle{spiebib} 

\end{document}